\documentclass[11pt]{article}

\usepackage{graphicx}
\usepackage{xcolor}
\usepackage[nodayofweek]{datetime} 
\usepackage{natbib}

\usepackage[letterpaper]{geometry} 
\usepackage{titling}
\usepackage{amsmath}
\usepackage{booktabs} 
\usepackage{array} 
\usepackage{multirow}

\usepackage{enumitem}
\setlist[enumerate,1]{wide}

\usepackage{setspace}
\usepackage{appendix}
\usepackage{hyperref}

\usepackage[explicit]{titlesec}
\titleformat{\section}{\large \bfseries \scshape}{\thesection.}{1ex}{#1}
\titlespacing{\section}{0pt}{1.5ex plus 0.5ex minus 0.5ex}{0.5ex plus 0.5ex minus 0.5ex}
\titleformat{\subsection}{\bfseries \itshape}{\thesubsection.}{1ex}{#1}
\titlespacing{\subsection}{0pt}{1ex plus 0.25ex minus 0.25ex}{0.0ex plus 0.25ex minus 0.25ex}
\titleformat{\subsubsection}[runin]{\bfseries}{\roman{subsubsection}}{1ex}{#1.}
\titlespacing{\subsubsection}{0pt}{1.5ex plus 0.5ex minus 0.25ex}{1em}

\newcommand{\der}{\mathop{}\!\mathrm{d}}

\DeclareMathOperator{\Var}{Var}
\DeclareMathOperator{\Cov}{Cov}

\newcommand{\q}{\mathbf{q}}
\newcommand{\avec}{\mathbf{\alpha}}
\newcommand{\W}{\mathbf{W}}
\newcommand{\N}{\mathbf{N}}
\newcommand{\cd}{\bullet}

\newcommand{\Fst}{F_{\mathrm{ST}}}
\newcommand{\aeq}{a^*}
\newcommand{\Sel}{\mathcal{S}}
\newcommand{\rr}{r}
\newcommand{\Rr}{r}
\newcommand{\Bb}{B}
\newcommand{\Cc}{C}

\newcommand{\ol}[1]{\overline{#1}}

\begin{document}

\begin{titlingpage}
\setlength{\droptitle}{2em}
\pretitle{\begin{center}\LARGE}
\posttitle{\par\end{center}\vskip 2.5em}

\title{ \bfseries Pathways to social evolution: \\
reciprocity, relatedness, and synergy}
\author{Jeremy Van Cleve$^{1}$ \and Erol Ak\c{c}ay$^{2}$}
\date{}
\maketitle

\vfill

\vspace{1em}

\begin{itemize}
\item[$^1$]
National Evolutionary Synthesis Center (NESCent) \\
2024 W. Main Street Suite A200 \\
Durham, NC 27705 USA\\
e-mail: \href{mailto:vancleve@nescent.org}{vancleve@nescent.org}
\item[$^2$] Department of Biology \\
University of Pennsylvania \\
433 S. University Avenue \\
Philadelphia PA, 19104 USA\\
e-mail: \href{mailto:eakcay@sas.upenn.edu}{eakcay@sas.upenn.edu}
\end{itemize}

\begin{flushright} \textit{Date modified: \today} \end{flushright}
\end{titlingpage}

\doublespace

\begin{abstract}

Many organisms live in populations structured by space and by class, exhibit plastic responses to their social partners, and are subject to non-additive ecological and fitness effects. Social evolution theory has long recognized that all of these factors can lead to different selection pressures but has only recently attempted to synthesize how these factors interact.  Using models for both discrete and continuous phenotypes, we show that analyzing these factors in a consistent framework reveals that they interact with one another in ways previously overlooked.  Specifically, behavioral responses (reciprocity), genetic relatedness, and synergy interact in non-trivial ways that cannot be easily captured by simple summary indices of assortment.  We demonstrate the importance of these interactions by showing how they have been neglected in previous synthetic models of social behavior both within and between species.  These interactions also affect the level of behavioral responses that can evolve in the long run; proximate biological mechanisms are evolutionarily stable when they generate enough responsiveness relative to the level of responsiveness that exactly balances the ecological costs and benefits.  Given the richness of social behavior across taxa, these interactions should be a boon for empirical research as they are likely crucial for describing the complex relationship linking ecology, demography, and social behavior.

\end{abstract}

\newpage

\section{Introduction}

Explaining the evolution of social behaviors has been a goal of evolutionary theory going back to Darwin's time. The modern theory for the origin of social behaviors started with the seminal work of \citet{Hamilton1964,Hamilton1964a}, which showed that genetic relatedness has a profound influence on evolutionary dynamics. Later work added other significant factors, starting with conditional and responsive behaviors \citep{Trivers1971, Axelrod1981} and extending to non-additive fitness interactions \citep{Queller1984,Queller1985}. Theorists analyzed many models that apply and extend these mechanisms \citep[e.g., extending reciprocity to include indirect responses,][]{Nowak1998}. At the same time, empirical researchers tested the role of various mechanisms developed in the theoretical literature \citep[e.g.,][]{Queller1998,Hughes2008,Schino2009}.

Whereas early discussions of social evolution usually pitted one mechanism against another, recent efforts are increasingly focused on synthesizing and integrating the various pathways of social evolution \citep[e.g.,][]{Marshall2003,Sachs2004,Lehmann2006,Nowak2006,West2007a,Fletcher2009,Queller2011}. This is perhaps the surest sign that social evolution theory has entered a mature phase. These efforts at synthesis are not always without controversy, but they have served to illuminate the connections between various models devised over the decades. In this paper, we aim to contribute to this synthesis by focusing specifically on the relation between three important concepts: assortment of genotypes (relatedness), behavioral responsiveness (e.g., reciprocity), and non-additive (e.g., synergistic) effects of social behaviors on fitness. These topics have of course been written about in the past; yet, how these concepts relate to one another is still not well understood. 

Two of the most fundamental mechanisms for the evolution of cooperation are responsive (and conditional) behaviors such as reciprocity, and genetic assortment, or relatedness, between individuals \citep{Lehmann2006}. Both factors are clearly in operation in many social interactions. Cooperative breeding, for example, almost always involves interactions with relatives of varying degree. At the same time, most cooperatively breeding species also exhibit conditional behavior; for example, non-helpers might get evicted from the breeding group \citep{Balshine-Earn1998}. Likewise, reciprocal food-sharing among vampire bats can occur amongst relatives as well as non-relatives \citep{Wilkinson1984}. 

Non-additive interactions between individual phenotypes also play a prominent role in social evolution. Sometimes, the non-additivity is a direct function of organismal metabolism, as in the case of the plant pathogen \textit{Agrobacterium tumefaciens}, which induces infected hosts to produce opines that can be used as a resource only by those cells that also carry the virulence plasmid \citep{White2007}. In other cases, the ecological context might create non-additive payoffs. For example, in the penduline tits (\textit{Remiz pendulous}), one or both of the parents desert the nest after eggs are laid. Data on breeding success suggests that the reason for this behavior is in part that a single parent has better than even chance of raising the clutch \citep{Szentirmai2007}. Hence, the second parent adds less to breeding success than the first parent, meaning a negative non-additivity. It has been argued that the opposite is true for many bird species with biparental care \citep{Wesolowski1994}.  It is worth noting that in both of these examples, as perhaps with most instances of non-additive interactions between phenotypes, the traits in question are plastic. For example, the \textit{Agrobacterium} pathways that induce opine production are not expressed constitutively but depend on quorum-sensing instead \citep{White2007}. Likewise, the male and female decisions to desert in penduline tits appear context-dependent; the same individual can care or desert according to factors such as brood size \citep{Persson1989,Valera1997}. 

Yet, despite copious theoretical work on the individual importance of conditional behaviors, genetic assortment, and non-additive payoffs, the interplay between these mechanisms remained relatively unexplored until recently \citep{Queller1985,Lehmann2006,Fletcher2006,Foster2006,Fletcher2009,McGlothlin2010,Queller2011,Akcay2012}. Much of this recent work neglects to systematically study the interactions between these factors \citep[with the notable exception of][]{Lehmann2006,McGlothlin2010,Akcay2012}, which we show can lead to an incomplete or incorrect view of how selection affects social behaviors.  Moreover, properly accounting for these interactions allows us to gain novel insights into when conditional behaviors should be evolutionarily stable and to empirically assess the level of responsiveness across demographically diverse species.

Our paper begins with two models: first, we use a discrete-action model to clarify the relationship between genetic assortment (e.g., relatedness), behavioral responses (a generic notion of reciprocity), and synergistic or non-additive interactions. Second, we generalize the results of the discrete-action model by presenting a population genetic model for the evolution of a continuous trait in a structured population, where the structure can be due to any combination of age, stage, sex, or spatial location.  Using these models, we show how previous analyses that do not systematically track the effect of behavioral responses \citep{Fletcher2009,Queller2011} have neglected a crucial interaction between responsiveness and relatedness \citep{McGlothlin2010,Akcay2012}.  Rather than combining additively in the equation for the effect of selection \citep{Queller2011}, relatedness and responsiveness interact through a product term that implies each of these assortment mechanisms contributes distinct effects \citep[contra][]{Fletcher2009}.  We also show this crucial interaction to be missing in models of interspecific mutualisms \citep{Foster2006}, which leads to difficulties in interpreting mechanisms that stabilize mutualisms.

More generally, our discrete model allows us to show how synergistic interactions can affect the evolutionary stability of social behaviors and even change the payoff structure of the game resulting in a different game entirely.  This is important because helping is more likely to evolve in Hawk-Dove \citep{Maynard-Smith1973} or Stag-Hunt \citep{Skyrms2001} games than in the Prisoner's Dilemma game.
Using our continuous-trait model, we show that proximate physiological or psychological mechanisms evolve in the long run only when they produce ``enough'' responsiveness relative to the responsiveness level that exactly balances the demographic costs and benefits of the behavior.  This evolutionarily stable level of responsiveness is a function of both the absolute level of relatedness and the interaction between relatedness and responsiveness.  Finally, we highlight in the discussion how recent work in quantitative genetics that measures relatedness and responsiveness requires a proper accounting of the interaction of these two mechanisms.

\section{Discrete action model}
\label{sec:recip}

We consider a 2x2 game played repeatedly between two individuals a large number of times. We take the game to be symmetric for ease of exposition, but this assumption can be relaxed with no consequence for our main argument. Label the two actions $C$ and $D$. Rather than focusing on a particular conditional strategy (e.g., the Tit-for-Tat \citep{Axelrod1981}), we model a wide range of propensities to respond to one's partner. We assume that each individual $i$ has two traits: $a_i$, which is the action ($C$ or $D$) that the individual $i$ intrinsically ``prefers'', and $\rho_i$, the probability that individual $i$ overrides its intrinsic preference and reciprocates its opponent's action in the last round. Hence, in common parlance, $a_i$ corresponds to whether individuals are cooperators or defectors, and $\rho_i$ denotes the degree to which individuals are reciprocators. Both traits are genetically encoded; we take $a_i$ to be binary and $\rho_i$ to be a continuous trait between $0$ and $1$.  A straightforward extension of this model could treat $a_i$ as a continuous trait by interpreting it as the intrinsic probability of playing $C$ vs.\ $D$.

Consider an interaction between an intrinsic cooperator and an intrinsic defector (i.e., $a_1=C$ and $a_2=D$). Suppose at stage $t$, player 1 (the $C$-type) played $C$ and player 2 (the $D$-type) also played $C$ (because he was reciprocating cooperation in the previous round). We label this situation as $(C,C)$, where the elements correspond to the first and second players' actions, respectively. Let us label the response coefficients of the two individuals as $\rho_c$ and $\rho_d$ in according with intrinsic preferences for $C$ and $D$, respectively. In that case, the probability that they will keep playing $(C,C)$ at stage $t+1$ is the probability that the $D$-type individual keeps reciprocating, i.e., $\rho_d$. Similarly, the probability that the $t+1$ stage will result in $(C,D)$ is $1-\rho_d$, as this is the probability that the $D$-type will revert to his intrinsic preference. Finally, the probabilities of playing $(D,C)$ and $(D,D)$ at $t+1$ given the actions at $t$ is zero, since the $C$-type player will always play $C$ regardless of whether she plays her intrinsic preference or reciprocates. In a similar way, we can work out the probabilities of playing any action pair at $t+1$ conditional on the action pair at $t$, which is summarized with the transition matrix $\mathbf{M} = [m_{ij}]$,
where $m_{ij}$ gives the probability of being in state $j$ at $t+1$ when the game was in state $i$ at $t$, and states are ordered as $\{(C,C), (C,D), (D,C), (D,D)\}$. So, $m_{11}=\rho_d$, $m_{12}=(1-\rho_d)$, etc.. We assume that the game is played for a very large number of periods with no discounting, so that the proportion of time each pair of actions is played is given by the stationary distribution of the Markov chain $\mathbf{M}$. Taking the dominant left-eigenvector of $\mathbf{M}$ and normalizing it, we find the stationary distribution of the game between a $C$-type and a $D$-type individual, $\mathbf{\pi}$, to be
\begin{equation}
\mathbf{\pi}=\left((1-\rho_c)^2 \rho_d \: , \: (1-\rho_c)(1-\rho_d)\: , \: (1-\rho_c)\rho_c(1-\rho_d)\rho_d\: , \: \rho_c (1-\rho_d)^2
\right) (1-\rho_c\rho_d)^{-2} \: .
\label{statdist}
\end{equation}
The expected payoff of a $C$-type player from playing with a $D$-type player is the fraction of time spent at each outcome (given by the elements of $\pi$) multiplied by the payoff to her from that outcome. Suppose that the stage game is an additive Prisoner's Dilemma: cooperation yields a benefit of $b$ to the partner, at a net cost of $c$ to the cooperator. Defection carries no cost and yields no benefit. Thus, the expected payoff to a $C$-type player when playing with a $D$-type, $w_{CD}$, is given by
\begin{align}
w_{CD}=\frac{(1-\rho_c)(b \rho_d-c)}{1-\rho_c\rho_d} \: .
\end{align}
With a similar calculation, we find the expected payoff to a $D$-type from playing with a $C$-type to be
\begin{align}
w_{DC}=\frac{(1-\rho_c)(b- c\rho_d )}{1-\rho_c\rho_d} \: .
\end{align}
The expected payoff to a $C$-type from playing with another $C$-type is trivial; since both players start with $(C,C)$ and keep playing these actions, $w_{CC}=b-c$. Similar reasoning yields $w_{DD}=0$. 

In order to determine which type of individual is preferred by natural selection, we calculate the condition when the frequency of intrinsic cooperators with responsiveness $\rho_c$ increases in a population with intrinsic defectors of responsiveness $\rho_d$. Suppose that some mechanism (e.g., population viscosity, kin-recognition, or greenbeards) causes cooperators to assort themselves so that an intrinsic cooperator interacts with another intrinsic cooperator with probability $\rr_c$; likewise, intrinsic defectors assort among themselves with probability $\rr_d$.  When not assorting specifically with their own types, individuals are paired with a partner randomly chosen from the population \citep[this is analogous to the classic assumption concerning populations with related individuals; see equation 1 in][]{Rousset2000}.  A full description of the dynamic of frequency change in the population would require a specification of the mechanism that generates assortment and how it depends on the frequency of genotypes. However, we can leave this unspecified if we only require a qualitative picture of frequency change over the course of a single generation.

Given our assumptions above and that the frequency of the intrinsic cooperative genotype is $q$, the expected fitnesses of an intrinsic cooperator and an intrinsic defector are, respectively,
\begin{equation}
\begin{aligned}
W_C&=(\rr_c + (1-\rr_c) q) \, w_{CC}+(1-\rr_c)(1-q) \, w_{CD} & \text{and} \\
W_D&=(1-\rr_d) q \, w_{DC}+ (\rr_d + (1-\rr_d)(1-q) \, w_{DD} \: .
\end{aligned}
\label{eq:fitness}
\end{equation}
The condition for an increase of intrinsic cooperators is $W_C - W_D > 0$, which leads to
\begin{align}
\frac{b}{c} & > \frac{1-(1-q) \rho _c - q \rho _d + (1-q) \rr_c \rho _c \left(1-\rho _d\right) + q \rr_d \left(1-\rho _c\right) \rho _d}{q \rho _c + (1-q) \rho _d + (1-q) \rr_c \left(1-\rho _d\right) + q \rr_d \left(1-\rho _c\right) - \rho _c \rho _d} \: .
\label{inc1}
\end{align}
Three special cases help illustrate the meaning of this equation. First, suppose individuals do not respond to each other, $\rho_c=\rho_d=0$, and that the assortment probabilities are equal, $\rr_c=\rr_d=\rr$.  In this case, condition \eqref{inc1} becomes $\frac{b}{c}>\frac{1}{\rr}$, which of course is Hamilton's rule \citep{Hamilton1964}.  Second, suppose both the intrinsic cooperators and defectors have the same responsiveness, $\rho_c=\rho_d=\rho$, and assortment is zero, $\rr_c=\rr_d=0$. Then, condition \eqref{inc1} simplifies to $\frac{b}{c}>\frac{1}{\rho}$, which is equivalent to other rules for the evolution of responsiveness \citep{Akcay2009} and reciprocity \citep{Lehmann2006,Andre2007}.  Finally, giving intrinsic cooperators and defectors the same assortment probability and responsiveness, $\rr_c=\rr_d=\rr$ and $\rho_c=\rho_d=\rho$, yields
\begin{align}
\frac{b}{c}>\frac{1+ \rr \rho}{\rr + \rho} \label{incsym} \: ,
\end{align}
which is analogous to equation (4) in \citet{Lehmann2006}.  In condition \eqref{incsym}, the assortment coefficient $\rr$ (which is now analogous to relatedness) and responsiveness $\rho$ have symmetric effects on the increase condition but remain distinct components because both the product and sum of $\rho$ and $\rr$ appear. 
Not coincidentally, expression \eqref{incsym} is identical in structure to the condition for an evolutionary increase in cooperation we have previously derived for continuous public goods games \citep{Akcay2012} and to other previously derived results \citep{Lehmann2006,McGlothlin2010}.

It is interesting to contrast our analysis with an earlier model by \citet{Marshall2003} that addresses a similar problem. In this model, five unlinked loci determine the initial strategy of an individual and the responses to each combination of strategies by the two players in the previous round. \citeauthor{Marshall2003} measure reciprocity with the locus that determines whether an individual retaliates by defecting after it cooperated and its opponent defected (a $(C,D)$ outcome in our notation). They find selection for retaliation decreases as relatedness increases. Later, \citet{Marshall2009} finds that this is due to negative non-additivity in the payoffs of the Prisoner's Dilemma game they use.  In contrast to these models, we model reciprocity as a generalized propensity to do what the opponent did in the last round regardless of what the focal individual did previously. As such, our results are not readily comparable, since in our model, the four conditional responses (one to each potential outcome) are inherently linked to each other. Thus while selection for retaliating after a $(C,D)$-round might decrease with increased relatedness in our model (because relatedness favors increased cooperation), this decrease is counteracted by increased selection to copy the opponent's behavior after a $(C,C)$ outcome. 

\section{Population genetic model}
\label{sec:popgen}

We now turn to a population genetic framework that will allow us to generalize the above results to interactions among an arbitrary number of individuals who may be from different demographic classes (e.g., males and females, queens and workers, etc.). A population genetic model will also allow us to gain insights into how the long-term evolutionary stability of social behaviors depends on the ecological costs and benefits and the proximate mechanisms of behavior. At the same time, such generality involves a significant increase in the complexity of the model, and hence a different set of simplifying assumptions are necessary. 

In particular, we are interested in determining when a mutant allele that codes for some phenotype with social effects will, on average, increase in frequency in a population and reach fixation.  The proper quantity to calculate in this case is the fixation probability of the mutant allele \citep{Rousset2003}.  For structured populations, the fixation probability may be impossible to obtain analytically, so we follow a standard approach that approximates that probability using a first-order Taylor-series expansion in $\delta$, which is the difference between the mutant and resident phenotypes \citep{Rousset2000,Rousset2004}.  Population genetics theory shows that the first-order term in the expansion is sufficient to calculate the standard conditions for the evolutionary stability of phenotypes \citep{Rousset2000,Rousset2004}.  Moreover, this first-order term is proportional to a quantity that is much easier to calculate called the ``selection gradient'', which is the derivative of the expected change in the frequency of the mutant allele with respect to $\delta$ or $\der \Delta q/\der \delta$.  More intuitively, we can think of $\Delta q \approx \left(\der \Delta q/\der \delta \right) \delta$ when $\delta$ is small.  This means that the sign of the selection gradient $\der \Delta q/\der \delta$ tells us when the mutant allele is expected to increase or decrease in frequency or, in terms of fixation probability, when the mutant allele is more or less likely than a neutral allele to fix in the population.

To obtain the selection gradient, $\der \Delta q/\der \delta$, we first write a general expression for $\Delta q$ in a class structured model using standard theory in population genetics and evolutionary demography \citep{Taylor1990,Charlesworth1994,Caswell2001,Rousset2004}.  We do this using an ``individually'' centered approach that was popularized in population genetics by the Price equation \citep{Price1972,Price1970}. The general expressions for $\Delta q$ and $\der \Delta q/\der \delta$ are derived in Appendix \ref{sec:app-class} in the Supporting Information (equations \ref{eq:deltap} and \ref{eq:1order}, respectively). While general, these expressions do not allow us to understand the interplay between behavioral and demographic mechanisms because we need to specify how the phenotype affects behavior and how behavior affects fitness. 

In order to simplify the analysis, we assume a relatively simple social interaction that is unaffected by the class structure (though class structure can still affect genetic relatedness).  Assume that each individual in each class participates in a single social interaction over the course of its lifetime with $n-1$ other individuals from the same class (e.g., members of a patch interact together).  Each individual chooses actions in this social interaction (e.g., an amount of effort in some helping behavior).  Over the course of the social interaction, individuals are assumed to interact with one another many times, and their actions are assumed to reach some equilibrium value \citep{Akcay2009,Akcay2012} due to a negotiation process \citep{McNamara1999,Andre2007,Akcay2009}. This equilibrium action is $\aeq_{k}$ for individual $k$ and could, for example, measure the amount of effort individual $k$ invests in helping its social partners at the conclusion of a negotiation over this amount.  If we perturb this equilibrium by changing the action of individual $k$, we can measure the effect of this perturbation on the action of individual $m$ by $\rho_{km}=\frac{\partial \aeq_{m}}{\partial \aeq_{k}}$, which is the response coefficient $\rho$ from section \ref{sec:recip} \citep{Akcay2009,Akcay2012}.

The negotiation process is driven by some proximate behavioral mechanism whose biological components (e.g., neurological structures) are genetically determined (whereas actions themselves are not inherited).  For simplicity, we assume that a single phenotype $p$ characterizes the proximate mechanism, and it is the evolution of this phenotype that we track with equation \eqref{eq:deltap}.  As an example, the phenotype could be a measure of hormone function in some brain region associated with cooperative behavior \citep[see][for a recent review]{Soares2010}.  The behavioral mechanism influences the negotiation of equilibrium actions, and we can measure this effect with $\frac{\partial \aeq_{k}}{\partial p_{m}}$, which is the effect of individual $m$'s phenotype on the action of individual $k$. This derivative could represent the effect of a change in hormone function in individual $m$ on the amount of effort individual $k$ invests in helping. The response coefficient can then be expressed as a function of these derivatives: $\rho_{km}=\frac{\partial \aeq_{m}}{\partial \aeq_{k}}=\left.\frac{\partial \aeq_{m}}{\partial p_{k}}\right/\frac{\partial \aeq_{k}}{\partial p_{k}}$.  Thus, response coefficients depend on the behavioral mechanism and the phenotypes of the individuals involved in the social interaction.

When the social interaction concludes, we assume that individuals obtain payoffs as a function of their equilibrium action and the equilibrium actions of their social partners.  We assume this payoff determines the fertility of each individual in the interaction (and affects no other fitness component); for example, an individual's fertility might be determined by the net calories it obtains from participating in group hunting activities where it expends calories when it helps the hunt and it obtains calories when prey are captured by itself or its partners.  An individual's fertility is denoted by $f(p_{k})$ for individual $k$. We can define the ``costs'' ($c$) and ``benefits'' ($b$) of the social interaction in terms of the affect of actions on fertility where $-c=\frac{\partial f}{\partial \aeq_{\cd}}$ and ``$\cd$'' represents the focal individual and $b=\frac{\partial f}{\partial \aeq_{\circ}}$ for some other individual ``$\circ$''.  Given these assumptions, we show in Appendix \ref{sec:app-class} in the Supporting Information that the selection gradient $\der \Delta q/\der \delta$ is proportional to
\begin{equation}
  \label{eq:1order:simp}
  \Sel = \underbrace{b \rho (n-1) - c}_{\text{``direct'' effect}} + \ \Rr \ \underbrace{(n-1) \left[ b \ (1 + \rho (n-2)) - \rho c \right]}_{\text{``indirect'' effect}} \: ,
\end{equation}
where $\rho$ is the response coefficient in a resident population monomorphic for the phenotype $p$ and $b$ and $c$ are evaluated at the resident phenotypic value $p$.  This result was previously derived by \citet{Akcay2012} and is analogous to expressions found in \citet{McGlothlin2010}.  The first term in \eqref{eq:1order:simp} is the ``direct'' effect on the fertility of a focal individual of its own phenotype and actions, whereas the second term is the ``indirect'' effect on the focal individual of the actions and phenotypes of its social partners. It is important to note that equation \eqref{eq:1order:simp} is essentially a rearrangement of Hamilton's classic rule \citep{Hamilton1964} where the effect of local competition due to kin is moved from the benefit and cost terms into the \textit{scaled} relatedness-coefficient $\Rr$ \citep{Queller1994,Rousset2004,Rousset2004a,Lehmann2010}. We derive an expression for $\Rr$ in Appendix \ref{sec:app-class} in the Supporting Information.  

Assuming that the phenotype $p$ corresponds to investment in a cooperative behavior in a Prisoner's Dilemma type of interaction, the condition for an increase in the level of cooperation derived from equation \eqref{eq:1order:simp} is $\Sel>0$ or
\begin{equation}
\label{eq:1order:bc}
\frac{b}{c} > \frac{1 + \Rr \rho (n - 1 )}{(n - 1) (\Rr + \rho + (n - 2) \Rr \rho)} \: .
\end{equation}
Condition \eqref{eq:1order:bc} is the $n$-player analog of \eqref{incsym}.  Even for arbitrarily large interaction groups ($n$ large), the symmetry between relatedness $\Rr$ and responsiveness $\rho$ is maintained.  This reinforces the notion that relatedness and responsiveness are distinct causal factors that interact synergistically and cannot be both replaced by a single index of ``assortment'' \citep{Fletcher2009}.  Moreover, social interactions with more than two individuals ($n>2$) introduce an additional product of relatedness and responsiveness; this product is scaled by $b (n-1) (n-2)$ and effectively accounts for the benefit a focal individual obtains when its relatives affect the behavior of a third party that interacts with the focal.

\begin{figure}
  \begin{center}
    \includegraphics[scale=0.65]{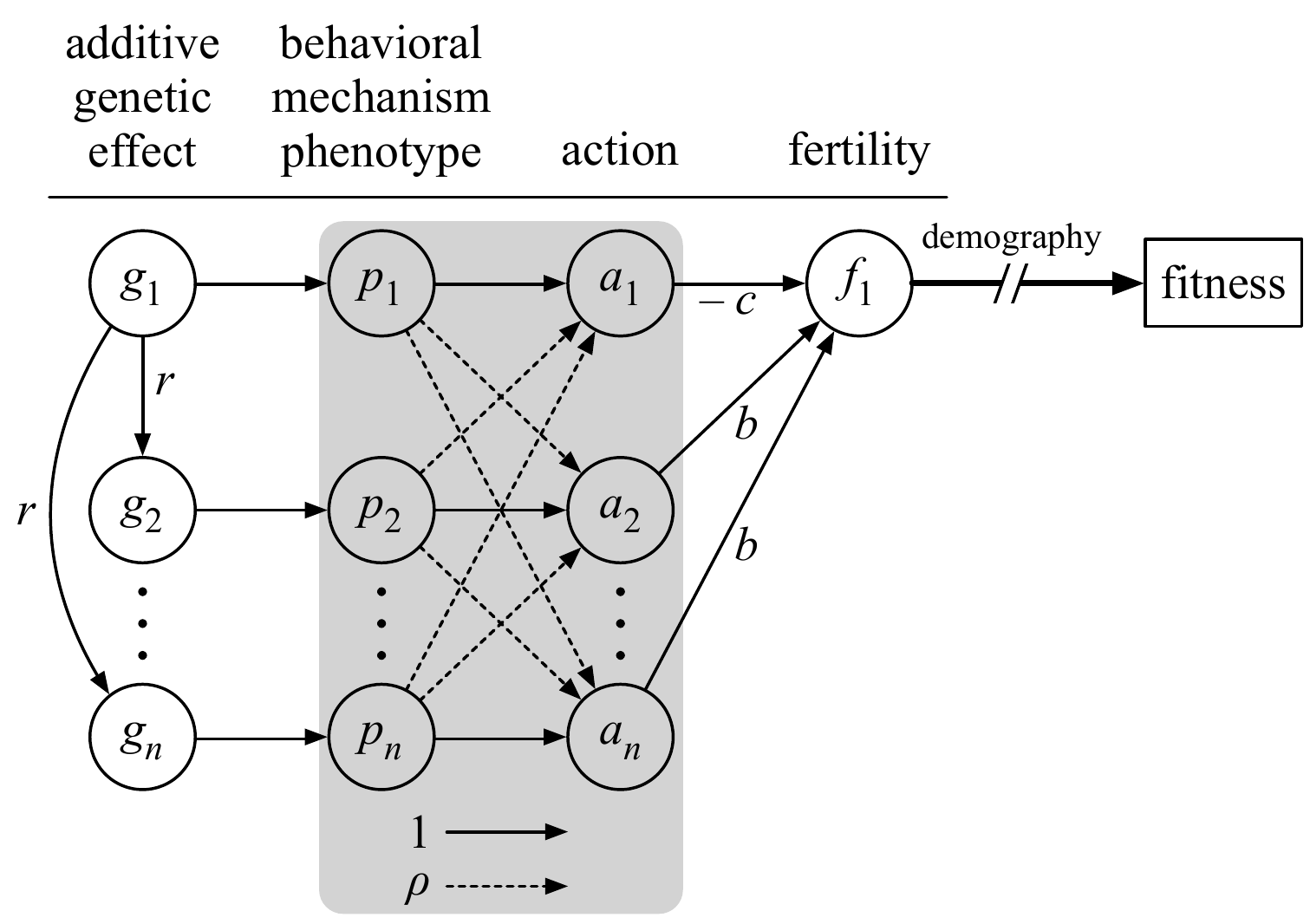}
    \caption{The path analysis diagram that describes the population genetic model in section \ref{sec:popgen} where individual $1$ interacts with its $n-1$ social partners.  Actions ($a$) are caused by the behavioral mechanism ($p$), which in turn is shaped by a heritable genetic component (additive genetic value, $g$). The path coefficient $\Rr$ measures genetic relatedness between the focal individual and its social partners. In the gray box, solid lines denote the effect that an individual's behavioral phenotype has on its own action and these paths have a coefficient of $1$ (by convention).  Dashed lines in the gray box have a coefficient of $\rho$ and represent the effect of an individual's phenotype on the actions of its social partners.  Actions affect the fertility of the focal individual where the focal decreases its own fertility by $c$ and its social partners increase it by $b$.  Finally, fertility is converted to fitness through demographic processes. The strength of selection on the phenotype $p$ is calculated by multiplying all the coefficients along a path and then summing all paths. This results in equation \eqref{eq:1order:simp} and shows how products of relatedness and responsiveness are created.}
    \label{pathdiagram}
  \end{center}
\end{figure}

Although we derive equations \eqref{eq:1order:simp} and \eqref{eq:1order:bc} from an explicit population genetic model, the same results can be obtained using the tools of path analysis \citep{Wright1920,Shipley2000}.  Briefly, path analysis in natural selection is a multiple regression tool used to measure the causal effect of genetics, environment influences, and other variables on fitness.  A path analysis diagram connects each causal variable to an effect variable via a directed edge.  The strength of such effects is measured by the path (or partial regression) coefficients that label each edge; unlabeled edges have coefficients that can be neglected since they are common to each path. Figure \ref{pathdiagram} shows the path analysis diagram for the population genetic model described in this section.

\section{Applications}

The discrete action and population genetics models developed above show that an interaction between responsiveness and relatedness emerges naturally when considering the effect of selection on a plastic social behavior.  Below, we illustrate the importance of the interaction between responsiveness and relatedness in four ways: first, we show that neglecting the interaction leads to incorrect predictions of the level of cooperation through the $b/c$ ratio and argue that past efforts to synthesize mechanisms for cooperation within species and for mutualisms between species have neglected this interaction; second, we show how the interaction between responsiveness and relatedness can be important in interspecific interactions, such as mutualisms; third, we extend the discrete action model to include non-additive or synergistic payoffs and demonstrate how responsiveness, relatedness, and synergy interact; and fourth, we use the population genetic model to derive a stability condition that reveals how proximate mechanisms evolve in the long run only when they generate enough responsiveness relative to the break even point determined by relatedness and the ecological costs and benefits of the behavior (i.e. $\Sel =0$).

\subsection{The interaction between reciprocity and relatedness}

The interaction between responsiveness (i.e., reciprocity) and relatedness in equation \eqref{eq:1order:simp} naturally breaks into two components: $- c (n-1) \Rr \rho$ represents the cost to a focal individual of increasing its own level of helping in response to the intrinsic increase in the level of helping of related group mates; and $b (n-1) (n-2) \Rr \rho$ represents the benefit the focal individual receives from third parties that increase their level of helping in response to the intrinsic increase in the level of helping of the focal individual's related group mates.  Both of these components are due essentially to the ability of individuals to respond to the actions of the focal's related group mates.  Neglecting these responses generates
\begin{equation}
\frac{b}{c} > \frac{1}{\Rr + \rho}
\label{eq:bc-no-rhor}
\end{equation}
as the condition for an increase in cooperation.  Condition \eqref{eq:bc-no-rhor}, which lacks the products of relatedness and responsiveness present in condition \eqref{eq:1order:bc}, effectively assumes that related group mates of the focal individual do not respond plastically to other individuals whereas unrelated individuals can respond plastically.  This assumption is very restrictive, which suggests that condition \eqref{eq:bc-no-rhor} is incorrect in most cases.

\begin{figure}
\small
  \begin{center}
    \includegraphics[scale=1]{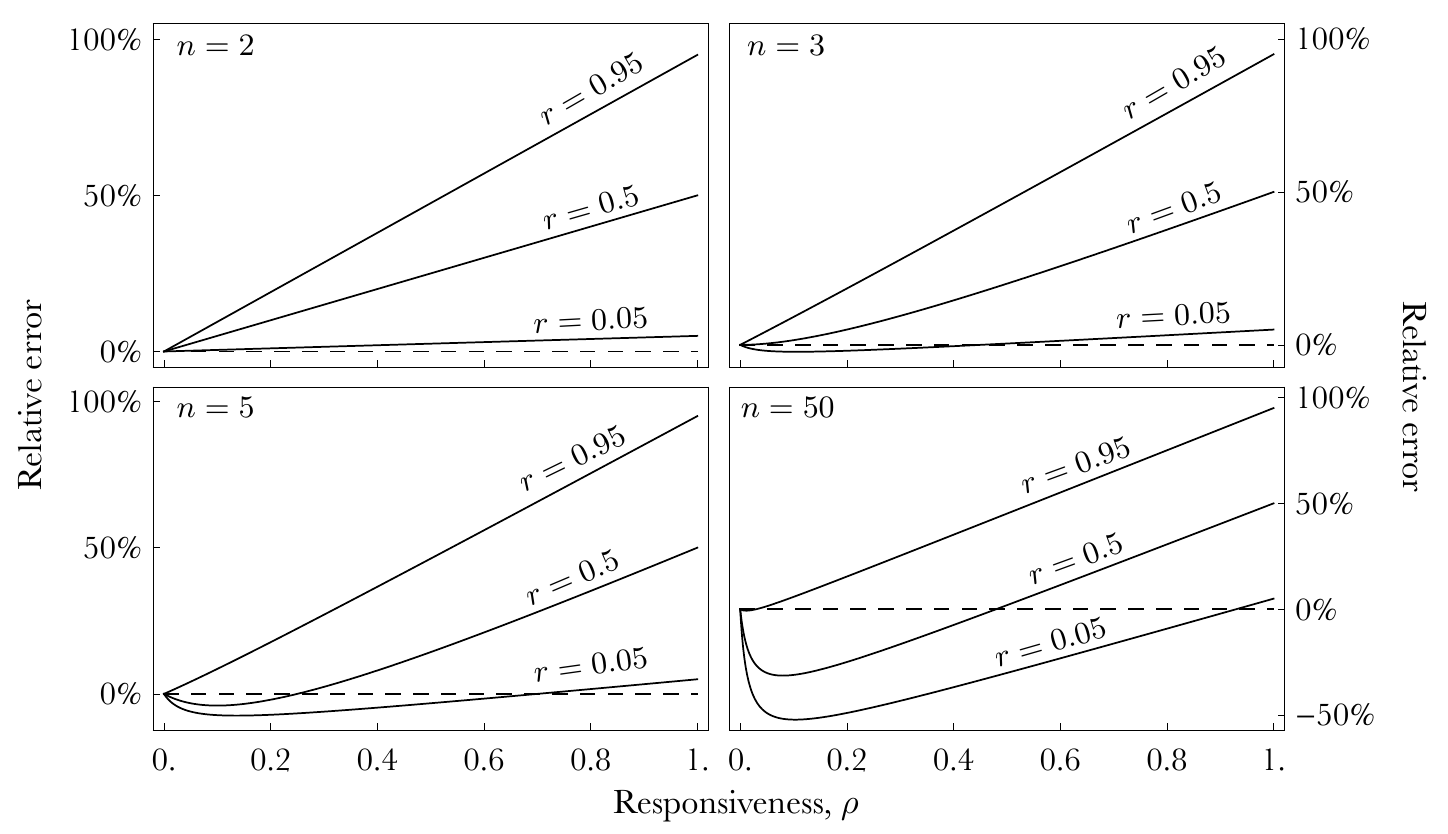}
    \caption{These panels plot the relative error of the $b/c$ expression given in \eqref{eq:bc-no-rhor} that lacks products of relatedness and responsiveness compared to the correct one given in \eqref{eq:1order:bc}.  Group sizes $n$ are given in each panel.  Each curve represents a different value of relatedness $r$ where the lower curve has $r=0.05$, the middle curve $r=0.5$, and the upper curve $r=0.95$.}
    \label{fig-rrho}
  \end{center}
\end{figure}

To visualize how neglecting the interaction between responsiveness and relatedness affects predicted levels of cooperation, we plot in Figure \ref{fig-rrho} the error of the incorrect expression in \eqref{eq:bc-no-rhor} relative to the correct $b/c$ expression in \eqref{eq:1order:bc} as a function of responsiveness.  For pairwise interactions ($n=2$), the relative error is always greater than zero regardless of the value of relatedness, which implies that neglecting the interaction between relatedness and responsiveness leads to an overestimate of the level of cooperation.  This is the case because the neglected interaction consists only of the cost the focal individual pays to respond to its relatives, $- c (n-1) \Rr \rho$.  For larger interaction groups ($n>2$), the interaction effect of third parties comes into play, $b (n-1) (n-2) \Rr \rho$.  This leads to an underestimate of the level of cooperation (relative error less than zero) for small values of responsiveness and relatedness and an overestimate for large values of either factor.  Thus, in precisely the situation where cooperation is expected to be low (both low relatedness and low responsiveness), the interaction between responsiveness and relatedness \textit{increases} cooperation.

Although an interaction between responsiveness and relatedness is a generic and important component of selection on social behavior as we have argued above, it can easily be neglected in efforts to synthesize different mechanisms of assortment when those efforts do not systematically track all of the effects of plastic responses.  Here, we give two examples of such efforts in an attempt to clarify how future syntheses can incorporate all the effects of different assortment mechanisms, whether they derive from behavioral plasticity within a generation or from demographic mechanisms operating across generations.  The first example is an ``expanded Hamilton's rule'' presented by \citet[equation 1]{Queller2011} as the following inequality
\begin{equation}
- c + \sum_i b_i \, r_i + \sum_i d_i \, s_i + \sum_i m_i \, f_i > 0 \: ,
\label{eq-queller-1}
\end{equation}
where $c$ is the cost of a cooperative act to the focal individual, $b_i$ is the benefit accruing to the $i$-th relative of the focal individual, $r_i$ the relatedness coefficient between the focal individual and the $i$-th relative. The third and fourth terms represent an expansion of the classical Hamilton's rule to include multiplicative effects and behavioral feedbacks, which Queller calls ``kind'' and ``kith'' selection, respectively. Of these terms, kind selection ($\sum_i d_i s_i$) involves a coefficient quantifying the deviation from additivity, $d_i$ for the $i$-th interaction partner, which is a ``synergistic'' fitness effect, and a synergy regression coefficient $s_i$. Finally, the last term ($\sum_i f_i m_i$) is the product of a coefficient measuring how the $i$-th social partner's phenotype responds to the focal individual's phenotype, $f_i$, and the fitness effect of this response $m_i$. Queller suggests that this expansion of the phenotypic version of Hamilton's rule represents more types of fitness effects in a causal manner than the classical form, and therefore might lead to better disentangling of the various pathways through which selection acts on social behaviors. While we agree with the basic premise, we show below that equation \eqref{eq-queller-1} lacks the interaction between relatedness and responsiveness and is therefore correct only under restrictive assumptions.

To see this, we simplify equation \eqref{eq-queller-1} by dropping the ``kind'' selection component (i.e. setting $d=0$) and obtain
\begin{equation}
- c + b \, \rr + m \, f > 0
\label{eq-queller-kinkith}
\end{equation}
for the case where there is a single class of individuals ($i=1$ in equation \eqref{eq-queller-1}).  In condition \eqref{eq-queller-kinkith}, the kith selection feedback coefficient, $f$, is analogous to our behavioral responsiveness measure $\rho$.  If we let the kith fitness effect $m$ be equal to the fitness effect for kin, $b$, then condition \eqref{eq-queller-kinkith} immediately becomes condition \eqref{eq:bc-no-rhor} that we evaluated above that lacks the interaction between responsiveness and relatedness.  The lack of this interaction stems from Queller's assumption, without any explicit justification, that the behavioral feedbacks and indirect fitness effects can be decomposed additively.  This assumption is correct only under very restrictive circumstances, namely when only individuals without relatedness to the focal individual respond to the focal individual and related individuals do not respond.  In Appendix \ref{sec:app-IGE} in the Supporting Information, we show that adopting this assumption explicitly in Queller's framework leads to condition \eqref{eq:bc-no-rhor}.  However, in most interactions among social partners with responsive phenotypes, both related and unrelated individuals will exhibit behavioral feedbacks, which suggests that a generalized Hamilton's rule that includes both kin and kith selection cannot be constructed by simply adding the kin and kith components together.

Our second example comes from \citet{Fletcher2009} who argue for the concept of ``interaction environments'' as a way to measure assortment genotype and behavior, regardless of the mechanism that generates the assortment.  \citeauthor{Fletcher2009} argue that the important quantity for determining whether cooperators proliferate more than defectors is the difference between the average interaction environment of cooperators, $e_c$, and that of defectors, $e_d$, which using our definitions of $b$ and $c$ yields
\begin{align}
\frac{b}{c} > \frac{1}{e_c-e_d} \: .
\label{incFD}
\end{align}
The interaction environments depend not only on the genetic assortment probabilities, $\rr_c$ and $\rr_d$, but also on how much each genotype ends up playing C and D against other intrinsic cooperators and defectors, respectively. For intrinsic cooperators, $e_c=\rr_c+(1-\rr_c)(q+(1-q)(\pi_1+\pi_3))$, where $\pi_i$ is the $i$th element of the stationary distribution of outcomes given in equation \eqref{statdist}, so the last term is simply the sum of the fraction of games between a $C$- and $D$-type where the outcome is $(C,C)$ and $(D,C)$ at the stationary distribution. Similarly, $e_d=(1-\rr_d)q(\pi_1+\pi_2)$. Substituting these expressions for $e_c$ and $e_d$ into condition \eqref{incFD} yields
\begin{align}
\frac{b}{c} > \frac{1- \rho_c \rho_d}{\rr_c (1-\rho_d)(1-q)+\rr_d q (1-\rho_c)+\rho_d(1-\rho_c)+q(\rho_c-\rho_d)} \:,
\end{align}
which simplifies to 
\begin{equation}
\frac{b}{c} > \frac{1+ \rho}{\rr + \rho}
\label{incFDsimp}
\end{equation}
when $\rho_c=\rho_d=\rho$ and $\rr_c = \rr_d = \rr$.  Comparing this expression with condition \eqref{incsym}, we can see that it is not the correct condition for an increase in cooperation as it lacks the interaction between relatedness and responsiveness in the numerator.  The reason the interaction environment method does not produce the correct interaction term $\rr \rho$ is because reciprocity and other types of behavioral responses affect a focal genotype's fitness not only through changing the social environment a genotype experiences, but also through changing the phenotype expressed by the focal genotype in response to this social environment.  Therefore, the implication that behavioral responses and genetic relatedness are only two special cases of the same fundamental assortment quantity does not hold in general since different assortment mechanisms can interact.

\subsection{Reciprocity and synergy}

In the above analyses, and in many biological models of social interactions, the marginal costs and benefits of cooperation and defection for the focal individual do not depend on the genotype of its partner.  However, it is often plausible that cooperators might obtain an extra benefit (or cost) from interacting with other cooperators, and this benefit (or cost) is called positive (or negative) synergy.  The importance of synergy, and non-additive interactions more generally, in understanding the evolution of social behavior was demonstrated by Queller in his seminal papers \citeyearpar{Queller1984,Queller1985}.  Since then, many studies have emphasized the positive and negative synergies can relax or constrain, respectively, the condition for cooperative behaviors to increase in frequency \citep{Queller1992a,Fletcher2006,Hauert2006,Lehmann2006,Lehmann2006a,Taylor2007,Ohtsuki2012,Taylor2012}.  If positive synergies are strong enough, they can also change the structure of the evolutionary game \citep{Fletcher2006,Hauert2006,Taylor2007}, which might turn a Prisoner's Dilemma game into a Stag-Hunt game.  

Since repeated interactions, such as reciprocity and behavioral responses, in a Prisoner's Dilemma game can also stabilize cooperation \citep{Axelrod1981,Fudenberg1986}, these interactions can also be viewed as changing the structure of the one-shot evolutionary game \citep{Fletcher2006,Lehmann2006,Andre2007,Taylor2007,Akcay2009,Akcay2011b}. In particular, \citet{Fletcher2006} show that reciprocity can alter the evolutionary game matrix to produce a game with non-additive payoffs. Going in the other direction, \citet{Marshall2003} and \cite{Marshall2009} study the evolution of behavioral responses in a particular non-additive game matrix. Notwithstanding these few studies, the interaction between behavioral responses and synergy has received no systematic study. Here, we show how behavioral responses and synergy have distinct effects on the dynamics of social traits and map the changes in the structure of the social game induced by both mechanisms simultaneously.

As is customary in models of synergy, we modify the Prisoner's Dilemma game from Section \ref{sec:recip} by adding a constant $d$ to the payoffs of both individuals when they both cooperate. When $d>0$, two cooperators produce more benefit together than twice the benefit of a single cooperator; when $d<0$, they produce less. These cases relate to cooperative acts by the partners being strategic complements versus substitutes, respectively, as defined in economics \citep{Bulow1985}.

Using the same methodology as in section \ref{sec:recip}, we calculate the expected payoff to an intrinsic cooperator when paired with an intrinsic defector and an intrinsic defector against an intrinsic cooperator as
\begin{equation}
\begin{aligned}
w_{CD}&=\frac{(1- \rho_c) \left( (b \rho_d - c)(1 - \rho_c \rho_d) + d \rho_d (1 - \rho_c) \right)}{(1 - \rho_c \rho_d)^2}  & \text{and} \\
w_{DC}&=\frac{(1- \rho_c) \left( (b - \rho_d c)(1 - \rho_c \rho_d) + d \rho_d (1 - \rho_c) \right)}{(1 - \rho_c \rho_d)^2} \: ,
\end{aligned}
\label{eq:payoffsyn}
\end{equation}
respectively.  These expressions indicate that reciprocity and payoff synergism can interact in complex ways. To better illustrate this interaction, suppose that both types of players have the same probability of reciprocation, $\rho_c=\rho_d=\rho$, and the same probability of assortment, $\rr_c=\rr_d=\rr$.  Then, as shown in Appendix \ref{sec:app-synergy} in the Supporting Information, the component that the payoff synergy, $d$, adds to the condition for cooperation to increase, $W_C-W_D>0$, is
\begin{equation}
d \left( p+\rr(1-p) + \frac{\rho (1-2p)(1-\rr)}{(1+\rho )^2} \right) \: .
\label{eq:synergy}
\end{equation}
Here, the first component, $p+\rr(1-p)$, matches previous work on synergy without reciprocity \citep{Wenseleers2006,Lehmann2006a}. The first thing to note from \eqref{eq:synergy} is that, compared with the non-synergistic case in condition \eqref{incsym}, synergy adds frequency dependence to the increase condition.  This is a well-known result \citep{Grafen1979,Queller1984,Lehmann2006a,Gardner2007,Ohtsuki2012,Taylor2012} that complicates the analysis of synergy.  Essentially, this frequency dependence arises from explicitly accounting for the frequency of different discrete pairings, which is necessary when payoffs are not additive.  Another way of thinking about this frequency dependence is that while relatedness or genetic assortment generally accounts for pairwise genetic correlations, accounting for synergy requires calculating triplet correlations, which are generally much more tedious to calculate \citep[for example]{Ohtsuki2012}.  Notably, this frequency dependence can be eliminated in certain stochastic models with symmetric mutation and weak selection \citep[e.g.,][]{Tarnita2009,Van-Cleve2013} even in the case of synergistic payoffs.

\begin{figure}
  \begin{center}
    \includegraphics[scale=1]{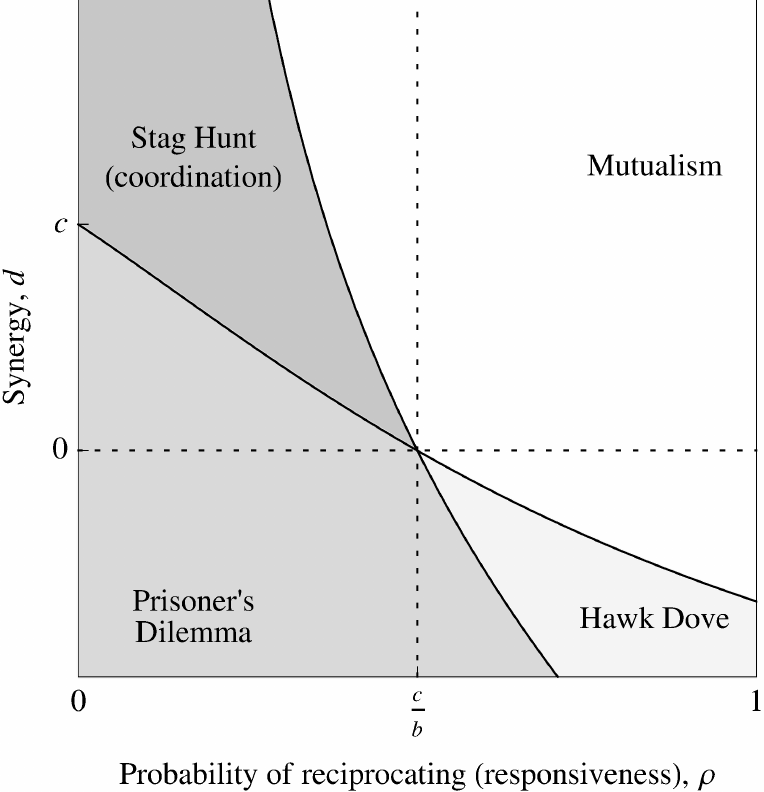}
    \caption{The different possibilities for how a Prisoner's Dilemma game with reciprocity and non-additive payoffs can get transformed, depending on the probabilities of reciprocation $\rho$ and the non-additive payoff or synergy component $d$.  Pure cooperation is unstable in the Prisoner's Dilemma, and pure defection is unstable in the Mutualism game. Both pure cooperation and defection are stable in the Stag-Hunt (or coordination) game.  Pure cooperation and defection are unstable in the Hawk-Dove game leaving an intermediate level of both stable (stable mixed strategy or polymorphism).}
    \label{synergy}
  \end{center}
\end{figure}

Second, the evolutionary game defined by the expected payoffs from the repeated game, $w_{CC}$, $w_{CD}$, $w_{DC}$, and $w_{DD}$, will take on different game structures depending on the values of $d$ and $\rho$. Figure \ref{synergy} illustrates the different possibilities. With $\rho<c/b$, a high enough synergism can transform the game from a Prisoner's Dilemma into a Stag-Hunt (coordination) game, which will have two evolutionarily stable strategies (ESSs): all cooperators or all defectors. If $\rho>0$, then still higher $d$ can turn the game into a Mutualism game, where intrinsic cooperators have an unconditional advantage. When $\rho>c/b$, the possibility of generating a coordination game disappears, and increasing $d$ turns the Prisoner's Dilemma into a Hawk-Dove game (anti-coordination or Snowdrift game) and eventually into a Mutualism game as $d$ increases further. To our knowledge, the interaction between reciprocity and synergy demonstrated in Figure \ref{synergy} has not been previously recognized. This interaction is particularly important in the light of recent results that show that non-additivity by itself does not change the conditions for the evolution of cooperation in populations with homogenous spatial structure \citep{Ohtsuki2012}, unless it changes the structure of the evolutionary game.

\subsection{The evolution of reciprocity}

While the discrete-action model allows us to derive intuitive relationships between relatedness, reciprocity, and synergy, it assumes a specific mechanism of repeated interactions that resembles tit-for-tat or direct reciprocity.  In contrast, the population genetic model in section \ref{sec:popgen}, where investments in cooperation are continuous, can handle families (in the mathematical sense) of flexible behavioral responses. In this section, we discuss how to analyze the evolutionary stability of such behavioral response families.

\begin{figure}
  \small
  \begin{center}
    \includegraphics[scale=1]{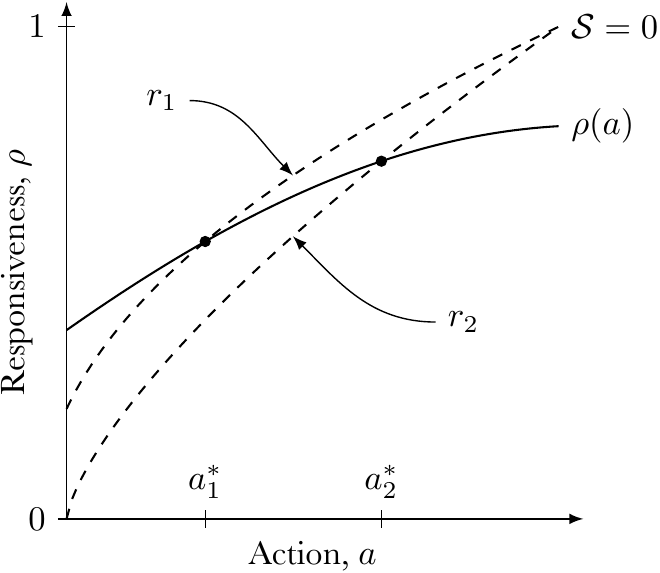}
    \caption{The graphical relationship between candidate evolutionarily stable response coefficients ($\rho$) as a function of individuals actions ($a$) and the response coefficients generated by a hypothetical proximate mechanism that we label $\rho(a)$. Here, actions might measure the level of helping, and the graphed proximate mechanism generates a positive relationship between responsiveness and actions as might be the case for individuals with an other-regarding preference for their partner's payoff. Candidate evolutionary stable values of $\rho$ (independent of any proximate mechanism) and $a$ solve $\Sel=0$ (see equation \ref{eq:1order:simp}) and are plotted as dashed curves for hypothetical values of $b$, $c$, and $n$ and for two hypothetical values of relatedness, $\Rr_1 < \Rr_2$.  The proximate mechanism above generates convergence stable actions, $a_1^*$ and $a_2^*$, since its response coefficient curve $\rho(a)$ crosses the $\Sel=0$ curves from above (see equation \ref{rhoCS}).}
    \label{fig-css}
  \end{center}
\end{figure}

In our population genetic approach, how individuals act and how responsive they are to their partners' actions are both functions of the phenotype ($p$) describing the proximate behavioral mechanism (see Figure \ref{pathdiagram}). Mechanistic examples of this phenotype $p$ could include the level of connectivity between particular brain regions or the sensitivity of particular neurons to neuromodulators such as oxytocin. Alternatively, one can think of $p$ at the functional level where it might represent, for example, the strength of an other-regarding preference individuals have towards their social partners \citep{Akcay2009}. Regardless of whether the phenotype $p$ is measured at the mechanistic or functional level, its role in our model is to provide a mechanism that connects how individuals act to how they respond to the actions of others.  Specifically, the phenotype (and the behavioral mechanism associated with it) implicitly defines a function $\rho(a)$ that measures the responsiveness level of an individual as a function of the action $a$ of its social partners when those partners are monomorphic for the phenotype. For instance, since more intense other-regarding preferences produce both higher investments in a group and more responsive individuals, $\rho(a)$ is an increasing function as depicted by the solid black curve in Figure \ref{fig-css}. 

Now, we can use the first order condition for evolutionary stability ($\Sel=0$ in \eqref{eq:1order:simp}), to solve for the responsiveness level $\rho$ that is necessary to make a given action $a$ evolutionarily stable at a given relatedness value $r$. In this way, the first order evolutionary stability condition $\Sel=0$ defines another set of curves (one for each value of relatedness) that maps actions, $a$, to responsiveness, $\rho$ (dashed curves in Figure \ref{fig-css}). Then, the intersection of the $\rho(a)$ curve obtained from the proximate mechanism and the curve obtained by setting $\Sel=0$ gives the evolutionarily stable value of the action and the phenotype that produces it. Note that different phenotypes for a single proximate behavioral mechanism or wholly different mechanisms will both in general generate different functions $\rho(a)$ and therefore different evolutionarily stable phenotypes and social behaviors.

Another important result our population genetic model yields concerns whether populations can evolve towards these evolutionarily stable phenotypes when they start from neighboring ones \citep{Eshel1981,Eshel1983,Taylor1989}.  This kind of evolutionary stability, called convergence stability, is often more important in the long-term evolution of continuous phenotypes than the traditional notion of strict evolutionary stability \citep{Wakano2012}.  A candidate phenotypic value is convergence stable when the derivative of $\der \Delta q/\der \delta$ with respect to the resident phenotypic value $p$ is negative \citep{Taylor1989}.  This condition amounts to
\begin{equation}
\label{eq:2order}
\frac{\der}{\der p} \left[ \frac{\der \Delta q}{\der \delta} \right]_{\Sel=0} \propto 
\left[ \frac{\der a}{\der p} \left(\frac{\der \rho}{\der a} \frac{\partial \Sel}{\partial \rho} + \frac{\der b}{\der a} \frac{\partial \Sel}{\partial b} 
+ \frac{\der c}{\der a} \frac{\partial \Sel}{\partial c} \right) \right]_{\Sel=0} \: < 0 \: ,
\end{equation}
where the derivatives in \eqref{eq:2order} are evaluated at $\Sel=0$ and measure how $\rho$, $b$, and $c$ change due to both direct and indirect changes in the phenotype $p$ and the resulting actions $a$.  Even in this generic form, condition \eqref{eq:2order} tells us that the evolutionary stability of a phenotype depends on how a change in the phenotype and the resulting action affects the level of responsiveness and how responsiveness changes selection for that phenotype.  In other words, it is not only the level of responsiveness that matters, but also how the underlying phenotype changes responsiveness and how responsiveness in turn alters the selection gradient.

More practically, the convergence stability condition can be interpreted graphically as in Figure \ref{fig-css}.  
With a little algebra and assuming $\frac{\der a}{\der p} > 0$ (a natural assumption if $p$ measures a cooperation phenotype and $a$ the level of investment in cooperation), condition \eqref{eq:2order} becomes
\begin{equation}
\label{rhoCS}
\left[ \frac{\der \rho(a)}{\der a} \right]_{\Sel=0} < \frac{\der \rho_{\Sel=0}}{\der a}
\end{equation}
where $\left[ \frac{\der \rho}{\der a} \right]_{\Sel=0}$ measures the slope of the behavioral mechanism $\rho(a)$ where it crosses the evolutionary stability curve ($\Sel=0$) and $\frac{\der \rho_{\Sel=0}}{\der a}$ is the slope of the evolutionary stability curve at that same point.  In other words, condition \eqref{rhoCS} implies that, in Figure \ref{fig-css}, the curve generated by the behavioral mechanism, $\rho(a)$, must cross the evolutionary stability curve, $\Sel=0$, from above.  Values of $\rho$ and $a$ on the $\Sel=0$ curve can be thought of as the actions that result from phenotypes that generate a level of responsiveness that exactly balances the costs and benefits of the behavior.  Thus, a behavioral mechanism that crosses the $\Sel=0$ curve from above is one that generates marginally more responsiveness than what is required to exactly balance the costs and benefits, and this is what makes it possible for evolution to converge to such a behavioral mechanism.  Importantly, the location of the $\Sel=0$ curve will depend on relatedness.  For Prisoner's Dilemma type interactions ($b(n-1)>c>0$), both $\partial \Sel / \partial \Rr >0$ and $\partial \Sel/\partial \rho >0$, which means that the $\Sel=0$ curve shifts down as relatedness increases.  In Figure \ref{fig-css}, we plot two $\Sel=0$ curves for hypothetical values of relatedness, $\Rr_1 < \Rr_2$, and its clear from the figure that increasing relatedness increases the convergence stable level of responsiveness for the proximate behavioral mechanism given by $\rho(a)$. 


\subsection{Interspecific mutualisms}

In Appendix \ref{sec:app-interspec}, we illustrate how the methodology of Section 3 and Appendix B in the Supporting Information can be applied to symbiosis (especially mutualistic ones) where host and symbiont phenotypes respond to each other and where both species exist in structured populations  \citep{Foster2006,Akcay2011,Akcay2014}. Our results show that both the host and symbiont population structure (characterized by within- and between-species genetic association coefficients) and phenotypic responses interact in previously unrecognized, non-additive ways \citep[compare for example, with][]{Foster2006}. This reinforces our main point, namely that phenotypic responses and genetic associations need to be considered in a consistent framework to be accurately integrated. In the interest of space, we refer the reader to Appendix \ref{sec:app-interspec} in the Supporting Information for the details of the model and the results. 

\section{Discussion}

One of the major goals of evolutionary and behavioral ecology is to elucidate the causal biological pathways that drive the evolution of social traits.  However, natural selection rarely operates through a single pathway, and therefore the various causal components of social evolution need to be integrated and their commonalities and interactions explored. Our main goal in this paper is to contribute to this synthesis effort. In so doing, we highlight that genetic assortment, behavioral responses, and non-additive interactions between phenotypes all have both distinct and interacting effects. Therefore, combining these components requires explicit and consistent modeling approaches. Using a model based on discrete phenotypes and a population-genetic model based on continuous phenotypes, we obtain the following results.

\begin{enumerate}[label=(\roman*)]

\item Phenotypic assortment is not a univariate quantity: it matters how much assortment comes from both behavioral responses and genetic assortment. Further, behavioral responses and genetic assortment cannot simply be added to each other, even when social interactions (payoffs) are additive. Rather, they interact with each other both additively and multiplicatively. Prior results analogous to this can be found in \citet{Lehmann2006}, \citet{McGlothlin2010} and \citet{Akcay2012}. Yet, despite this interaction being known for some time, its crucial role seems to be not widely appreciated. \citet{Akcay2012} is the first paper to our knowledge to highlight the synergistic and symmetric effect of genetic assortment and behavioral responses on the response to selection of social phenotypes in social groups of any size. In that work, we demonstrated how synergism between genetic assortment and behavioral responses can facilitate the transfer of fitness effects from the within-group level to the between-group level, a crucial step in evolutionary transitions in individuality \citep{Maynard-Smith1995,Michod2005}. These previous results assumed continuous phenotypes and within-species interactions; here we show that a similar synergism between genetic assortment and behavioral responses is found for discrete phenotypes and is important for species interactions as well.  

\item Non-additive or synergistic interactions can have important effects on whether cooperative behaviors increase in frequency in a population and whether these effects depend on behavioral responses and reciprocity.  Specifically, synergy and reciprocity can both change the structure of the evolutionary game \citep{Taylor2007}.  When this game starts as a Prisoner's Dilemma where cooperation is unconditional, increasing synergy or responsiveness levels can stabilize some level of cooperation.  However, whether all individuals cooperate fully depends on the combination of synergy and responsiveness that determines whether the game becomes a Stag-Hunt, Mutualism, or Hawk-Dove game (cf. \citealp{Van-Cleve2013}).

Even though synergistic interactions are easy to study using a discrete phenotypic model, the population-genetic model with continuous phenotypes shows the importance of non-additive payoffs or fitness in more subtle way.  These interactions will affect the evolutionary (convergence) stability of a phenotype and how the stable phenotype changes with underlying parameters. The former effect can be seen in equation \eqref{eq:2order}, which shows that non-additivity in the benefits and costs of a social phenotype (behavior) appear in the convergence stability condition (since $b$ and $c$ are themselves already first-order derivatives of fertility).  Non-additivity in benefits and costs relate directly to ecological mechanisms that convert the effort of individuals in the social interaction into payoffs in terms of fertility or survival.  For example, individuals who put effort into a group hunt \citep{Skyrms2001} may receive more return in terms of the amount of prey item per unit of caloric investment when more individuals join the hunt.  This kind of ecological synergy, or ``complementarity'' in economics \citep{Bulow1985}, can have large effects on the level of investment in cooperative interactions both within groups \citep{Akcay2009} and in structured populations \citep{Akcay2012,Cornforth2012}.

\item There is a simple graphical relationship required for specific proximate behavioral mechanisms (e.g., social preferences, social norms, punishment, etc.) to be evolutionarily stable in the long run.  When plotting the curve of responsiveness as a function of the evolving phenotype for the mechanism (level of other regard, strength of conformity, probability of punishment, etc), this curve must cross the curve that balances the costs and benefits of the behavior from above.  Moreover, since different levels of relatedness result in different curves that balance benefit and cost, what level of responsiveness evolves for a specific proximate mechanism can also be determined graphically.  

\end{enumerate}

\subsection{Empirical issues}

The empirical study of the interplay between behavioral responses and genetic assortment and the role of synergistic payoffs is still relatively new.  Most of the recent work linking behavioral responses and genetic assortment comes from the application of indirect genetic effect (IGE) models to laboratory, agricultural, and field-based populations \citep[e.g.][]{Moore2002,Petfield2005,Mutic2007,Ellen2008,Danielson-Francois2009,Bleakley2009,Teplitsky2010,Frere2010,Wilson2011,Hamilton2012}.  IGE models \citep{Moore1997,Wolf1999,McGlothlin2010} extend classical quantitative genetics \citep{Lynch1998} to include a coefficient, $\psi$, that measures the effect of a social partner's phenotype on the phenotype of a focal individual.  Importantly, the IGE coefficient $\psi$ directly maps onto our behavioral responsiveness coefficient $\rho$ \citep{Akcay2012} and is thus also a measure of reciprocity.   We know from equation \eqref{eq:1order:simp} (and analogously equation 18 in \citealt{McGlothlin2010}) that an interaction between behavioral responses and relatedness is important for selection on social traits, even when no such interaction is assumed to affect individual phenotypes (see equation \ref{eq-p} in the Supporting Information or equation 5 in \citep{McGlothlin2010}).   However, most of the empirical work to date spends little time on the interaction of behavioral responses and relatedness with the notable exceptions of \citet{Bleakley2009} and \citet{Frere2010}.  Moreover, some empirical work assumes an explicit interaction term in the regression model for individual phenotypes, and it is not yet clear how empirical estimates of this interaction can be tied to the predicted interaction term that arises in measures of selection on the trait.  Thus, more work needs to be done understanding the interaction between behavioral responses and relatedness both theoretically and empirically.

Empirical studies explicitly focused on non-additive payoffs are uncommon \citep[are notable exceptions]{Queller2003,Gore2009}. This paucity of studies does not reflect the importance of non-additive interactions in nature. Instead, non-additive and non-linear interactions are likely to be abundant and their effects strong.  For example, in parental care, many arguments implicitly or explicitly rely on the contributions of each parent combining non-additively \citep[e.g.,][]{Winkler1987,Akcay2009a}. We suggest that measuring non-additive interactions at both the payoff (fertility or survival) and fitness levels will prove important for explaining the evolution of social behaviors, regardless of whether the phenotypes in question vary continuously or are discrete characters with large effects.  

To conclude, a careful analysis of behavioral responses, genetic assortment, and non-additive interactions shows a complex interaction between these three pathways. We believe that focusing on this complexity will ultimately deliver a more complete and nuanced understanding of the evolutionary forces shaping social behaviors. The necessary mathematical framework for such understanding exists, though as we show in this paper, it needs to be applied with some care.  Once crafted, these theories can be tested against the growing abundance of sophisticated datasets that measure both networks of social interactions and important demographic and fitness-related variables \citep[e.g.][]{Holekamp2012,Apicella2012}.

\section*{Acknowledgements}

J.V. was supported by a Santa Fe Institute Omidyar Fellowship and the National Evolutionary Synthesis Center (NESCent), NSF \#EF-0423641.  E.A. was supported by National Science Foundation Grant EF-1137894, to S.A. Levin.


\clearpage
\bibliographystyle{evolution}
\bibliography{refs}

\newpage

\appendix

\titleformat{\section}{\normalfont\large\bfseries}{Appendix \thesection:}{1ex}{#1}
\numberwithin{equation}{section}

\section{Two-player game with synergy}
\label{sec:app-synergy}

We follow the discrete-action model defined in section \ref{sec:recip}.  Individuals are either intrinsic cooperators or intrinsic defectors who can either play their intrinsic strategy or reciprocate their partner's last action. Let $\rho_c$ and $\rho_d$ be the probabilities that an intrinsic cooperator and defector, respectively, reciprocate or copy their partner's last action.  We allow for non-random interactions among intrinsic cooperators and intrinsic defectors that might result from population structure or other assortation mechanisms and define $\rr_c$ and $\rr_d$ to be the probabilities that intrinsic cooperators and defectors, respectively, interact among themselves non-randomly.  Two cooperators who play each other each earn payoff $b-c+d$, where $b$ is the benefit, $c$ is the cost of cooperation, and $d$ is a synergy term that accounts for non-additive payoffs.  A cooperator and defector playing together results in a payoff of $-c$ for the cooperator and $b$ for the defector.  Two defectors playing each other each earn zero payoff.  Given this setup and following the analysis of section \ref{sec:recip}, the payoffs to an intrinsic cooperator paired with an intrinsic defector and vice-versa are
\begin{equation*}
\begin{aligned}
w_{CD}&=\frac{(1- \rho_c) \left( (b \rho_d - c)(1 - \rho_c \rho_d) + d \rho_d (1 - \rho_c) \right)}{(1 - \rho_c \rho_d)^2}  & \text{and} \\
w_{DC}&=\frac{(1- \rho_c) \left( (b - \rho_d c)(1 - \rho_c \rho_d) + d \rho_d (1 - \rho_c) \right)}{(1 - \rho_c \rho_d)^2}  \: ,
\end{aligned}
\end{equation*}
respectively, which are also given in equation \eqref{eq:payoffsyn} in the main text.  

We can plug the above values for $w_{CD}$ and $w_{DC}$, along with $w_{CC} = b - c + d$ and $w_{DD} = 0$, into the expression for fitness in equation \eqref{eq:fitness} to derive the condition for cooperation to increase, $W_C - W_D > 0$, which simplifies to
\begin{equation}
  \begin{aligned}
    b & \left( \frac{p \rho_c+(1-p)\rho_d+\rr_c (1-p)(1-\rho_d)+\rr_d p (1-\rho_c)-\rho_c \rho_d}{1-\rho_c \rho_d} \right)\\
    -c & \left(\frac{1-(1-p)\rho_c-p \rho_d+(1-p)\rr_c \rho_c (1-\rho_d)+\rr_d p \rho_d(1-\rho_c )}{1-\rho_c \rho_d}\right) \\
    +d & \left(\frac{(p+\rr_c (1-p)) (1-\rho_d) \left(1-\rho_c^2 \rho_d\right)+(1-p+p \rr_d) (1-\rho_c)^2 \rho_d}{(1-\rho_c \rho_d)^2} \right) > 0 \: .
  \end{aligned}
  \label{eq:app:incr}
\end{equation}
This expression simplifies to condition \eqref{inc1} when $d=0$.  Another simplification is to assume that the reciprocation and assortment probabilities are the same for cooperators and defectors, namely $\rho_c=\rho_d=\rho$ and $\rr_c=\rr_d=\rr$.  In this case, expression \eqref{eq:app:incr} simplifies to
\begin{equation}
b \frac{\rr+\rho}{1+\rho }- c\frac{1 + \rr \rho}{1+\rho } + d \left( (p+\rr(1-p)) + \frac{\rho (1-2p)(1-\rr)}{(1+\rho )^2} \right) > 0\: ,
  \label{eq:app:incr-simp}  
\end{equation}
which in turn simplifies to condition \eqref{incsym} when $d=0$.

\clearpage
\section{Allele frequency change in a class-structured model}
\label{sec:app-class}

Suppose that there are $K$ classes in the population, where an individual $k$'s class could represent the local population it lives in, its age or developmental stage, its sex, or some other phenotypic feature.  Each class contains $N_k$ individuals and the total population size is $N=\sum_k N_k$.  In order to track allele frequencies in different classes, we let $\q = (\ldots, q_{jk}, \ldots)$ be a column vector of length $N$ where $q_{jk}$ is the frequency of the allele in individual $k$ in class $j$.  Average allele frequencies in each class in the next time period are calculated by multiplying a transition matrix $\W = [w_{ijk}]$ times $\q$, where $w_{ijk}$ is the probability that an allele in a random individual in class $i$ descended from individual $k$ in class $j$.  The transition matrix is a function of the fitnesses of each class and the transmission processes, such as segregation, mutation, and migration, that affect how alleles move between classes.  Fitness can include both fertility and survival, and thus populations with both overlapping and nonoverlapping generations are possible. We assume that the fitness of individual $k$ in class $j$ is function of its phenotype, $p_{jk}$, and potentially the phenotypes of its social partners.  In the simplest cases, phenotype depends on genotype in a linear way where $p_{jk} = p + q_{jk} \delta$ and $\delta$ is a measure of how much the mutant differs from the resident phenotype $p$.  For non-additive genetic interactions (such as dominance or epistasis), phenotype depends on genotype in a nonlinear way \citep{Gardner2007}.

Calculating $\Delta q$ requires us to average the class frequencies in such a way that the average frequency remains constant in the absence of natural selection.  The appropriate weights for this average are the reproductive values of each class \citep{Fisher1930,Leslie1948,Taylor1990}, $\avec = (\ldots, \alpha_{k}, \ldots)$.  Thus, the change in the reproductive value weighted allele frequency is
\begin{equation}
  \label{eq:deltap}
  \Delta q = \avec \cdot \W \cdot \q - \avec \cdot \N \cdot  \q
\end{equation}
where $\N$ is a $K \times N$ matrix used to create class averages of the allele frequencies $\q$ and whose $k$-th row has the value $1/N_k$ in the columns $1+\sum_{l=1}^{k-1} N_l$ to $\sum_{l=1}^{k} N_l$ and zero elsewhere for all $k$.  Using equation \eqref{eq:deltap}, the derivative of $\Delta q$ with respect to $\delta$ is given by
\begin{equation}
  \label{eq:1order}
  \frac{\der \Delta q}{\der \delta} = \avec \cdot \frac{\der \W}{\der \delta} \cdot \q = \sum_{i,j,l=1}^{K} \sum_{k=1}^{N_j} \sum_{m=1}^{N_l} \alpha_i \frac{\partial w_{ijk}}{\partial p_{lm}} q_{jk} q_{lm}
\end{equation}
The components of equation \eqref{eq:1order} are reproductive value, $\alpha_i$, the effect of one individual's phenotype on another individual's probability of transmitting an allele from one generation to the next, $\frac{\partial w_{ijk}}{\partial p_{lm}}$, and the probability that those two individuals both have the mutant allele, $q_{jk} q_{lm}$.  When this last component, $q_{jk} q_{lm}$, is averaged over all individuals in a class, we obtain a genetic identity probability that can be related to identity by descent probabilities from classic population genetics \citep{Rousset2000,Rousset2004,Rousset2004a} that emerge as more common quantities such as $\Fst$ \citep{Wright1949} in simple models of population subdivision \citep{Rousset2004a} such as the island model \citep{Wright1931}.  Moreover, these genetic identity probabilities correspond to the genetic covariance terms found in regression models in social evolution \citep{Frank1997}, which when normalized by genetic variance correspond to the regression definition of relatedness \citep{Hamilton1972,Queller1992}.

Recall that the Taylor series is expanded around $\delta=0$, so the partial derivatives $\frac{\partial w_{ijk}}{\partial p_{lm}}$ in \eqref{eq:1order} are evaluated at neutrality where natural selection on fertility differences is absent.  Additionally, $\delta=0$ implies that all individuals in a class are equivalent, which means that the partial derivatives above fall into three cases: the first case is the effect of a focal individual's phenotype on its own probability of sending an allele to class $j$ in the next generation from its own class $i$, $\frac{\partial w_{ij}}{\partial p_{j \cd}}$; the second is the effect on that probability of another individual's phenotype from the same class as the focal, $\frac{\partial w_{ij}}{\partial p_{j \circ}}$; and the third is the effect on that probability of the phenotype of an individual in a different class than the focal individual, $\frac{\partial w_{ij}}{\partial p_{l}}$ where $l \neq j$.  Substituting these derivatives into \eqref{eq:1order}, we obtain
\begin{equation}
  \avec \cdot \frac{\der \W}{\der \delta} \cdot \q = \sum_{i,j=1}^{K} 
  \alpha_i N_j \left( \frac{\partial w_{ij}}{\partial p_{j \cd}} \ol{q_{j \cd}^2} 
    + (N_j-1) \frac{\partial w_{ij}}{\partial p_{j \circ}} \ol{q_{j \cd} q_{j \circ}} 
    + \sum_{l \neq j} N_l \frac{\partial w_{ij}}{\partial p_{l}} \ol{q_{j} q_{l}} \right) \: .
  \label{eq:1order-simp1}
\end{equation}

Further, we assume that the phenotype only affects fertility $f$ (though it could affect survival or demographic factors such as carrying capacity or dispersal rates), which means the derivatives in the right-hand side of \eqref{eq:1order-simp1} can be split into two components, the effect of the phenotype on fertility and the effect of fertility on the probability an allele in class $i$ descends from class $j$, $w_{ij}$.  The latter derivatives we represent with $\tau$ where $\tau_{ij,j\cd}=\frac{\partial w_{ij}}{\partial f_{ij \cd}}$, $\tau_{ij,j\circ}=\frac{\partial w_{ij}}{\partial f_{ij \circ}}$, and $\tau_{ij,l}=\frac{\partial w_{ij}}{\partial f_{il}}$ for $l \neq j$.  These derivatives essentially contain transmission probabilities due to dispersal, mutation, allelic segregation, age-dependent mortality, etc, and we assume they are independent of phenotype. Given these two classes of derivatives, 
\begin{equation}
  \frac{\partial w_{ijk}}{\partial p_{lm}} =
  \begin{cases}
    \frac{\partial w_{ij}}{\partial p_{j \cd}} = \tau_{ij,j\cd} \frac{\partial f_{ij}}{\partial p_{j \cd}} + (N_j-1) \tau_{ij,j\circ} \frac{\partial  f_{ij}}{\partial p_{j \circ}} 
    & \text{for $(l,m) = (j,k)$} \\
    \frac{\partial w_{ij}}{\partial p_{j \circ}} = \tau_{ij,j\cd} \frac{\partial f_{ij}}{\partial p_{j \circ}} + \tau_{ij,j\circ} \left( \frac{\partial  f_{ij}}{\partial p_{j \cd}} + (N_j - 2) \frac{\partial  f_{ij}}{\partial p_{j \circ}} \right)
    & \text{for $l=j$ and $m \neq k$} \\
    \frac{\partial w_{ij}}{\partial p_{l}} = \tau_{ij,l} \left( \frac{\partial f_{il}}{\partial p_{l \cd}} + (N_l-1) \frac{\partial f_{il}}{\partial p_{l \circ}} \right) & \text{for $l \neq j$} \: .
  \end{cases}
\label{eq:dwdp}
\end{equation}

Finally, we assume that the effect of phenotype on fertility is independent of the source or destination classes of the offspring; that is, we can define $-\Cc = \frac{\partial f_{ij}}{\partial p_{j \cd}}=\frac{\partial f}{\partial p_{\cd}}$ and $\Bb = \frac{\partial f_{ij}}{\partial p_{j \circ}}=\frac{\partial f}{\partial p_{\circ}}$.  Combining $\Cc$ and $\Bb$ with \eqref{eq:dwdp} into equation \eqref{eq:1order-simp1} results in
\begin{equation}
  \begin{aligned}
    \avec \cdot \frac{\der \W}{\der \delta} \cdot \q = & - \Cc \left[ \sum_{i,j=1}^{K} \alpha_i N_j \left( \tau_{ij,j\cd} \ol{q_{j \cd}^2} 
        + (N_j-1) \tau_{ij,j\circ} \ol{q_{j \cd} q_{j \circ}} + \sum_{l \neq j} N_l \tau_{ij,l} \ol{q_{j} q_{l}} \right) \right] \\
    &\hphantom{-} \Bb \left[ \sum_{i,j=1}^{K} \alpha_i N_j (N_j-1) \left( \tau_{ij,j\cd} \ol{q_{j \cd}^2} 
        + \left( \tau_{ij,j\cd} + (N_j-2) \tau_{ij,j\circ} \right) \ol{q_{j \cd} q_{j \circ}} \right) + N_j \sum_{l \neq j} N_l (N_l-1) \tau_{ij,l} \ol{q_{j} q_{l}} \right] \\
= & K_{\Cc} \left( - \Cc + \Rr \Bb \right) \: ,
  \end{aligned}
  \label{eq:1order-simp2}
\end{equation}
where $K_{\Cc}$ and $K_{\Bb}$ equal the first and second bracket in \eqref{eq:1order-simp2}, respectively.  Scaled relatedness is given $\Rr = K_{\Bb}/K_{\Cc}$.  The terms in parentheses on the last line of \eqref{eq:1order-simp2} are the same as the right-hand side of equation \eqref{eq:1order:simp}, once $\Cc$ and $\Bb$ are evaluated following the assumptions in that section that the phenotype $p$ is a parameter of some physiological/psychological mechanism that affects the underlying actions individuals choose, $a^*$, and the marginal effect of the action of one individual on the action of another is $\rho$.  A detailed exposition of this analysis is given by \citet{Akcay2012} where $-\Cc$ corresponds to $\frac{\partial F_i}{\partial p_i}$ in equation (2) and $\Bb$ to $\frac{\partial F_i}{\partial p_j}$ in equation (3) in that paper.  In most biological cases of interest, $K_{\Cc}>0$, which results in the proportionality in equation \eqref{eq:1order:simp}.  The convergence stability condition in \eqref{eq:2order} can be derived from equation \eqref{eq:1order:simp} simply by recalling the assumption that the phenotype affects fertility but not demographic quantities contained in $\Rr$.

\clearpage
\section{Uni-directional and bi-directional reciprocity in indirect genetic effects models}
\label{sec:app-IGE}

In this section, we will show that: (i) assuming a uni-directional type of reciprocity where the focal individual can affect its partner's phenotype, but not vice-versa, can lead to the expanded Hamilton's rule presented by \citet[equation 1]{Queller2011}; and (ii) bi-directional reciprocity leads to condition \eqref{incsym}, which is equivalent to results derived by us \citep{Akcay2012} and others \citep{Lehmann2006,McGlothlin2010}.  Notation in the following equations is taken from \citet{Queller2011} and generally follows quantitative genetic approaches to social evolution \citep[e.g.,][]{Queller1992a,Frank1997,Queller2011} with a particular focus on the methods in indirect genetic effect (IGE) models \citep{Moore1997,Wolf1999,McGlothlin2010}.


We begin with the phenotypic version of Hamilton's rule, derived from the Price equation \citep{Price1970}, that is equation (7) in \citet{Queller2011}, which is
\begin{equation}
\beta _{W P\cdot P'} + \beta _{W P'\cdot P} \frac{\Cov\left[G, P' \right]}{\Cov[G,P]} \: .
\label{eq-queller-7}
\end{equation}
where we use primes to denote variables associated with the social partner.  Suppose that the phenotype of the social partner is defined as 
\begin{equation}
P' = G' + \beta _{P' P} P + \epsilon' \: ,
\label{eq-pprime}
\end{equation}
which says that the phenotype of the social partner is a linear function of its average breeding value ($G'$), the effect of the phenotype of the focal ($\beta _{P' P} P$), and a random component with mean zero ($\epsilon'$, a standard assumption in quantitative genetics; see \citealp{Lynch1998}).  Suppose also that the phenotype of focal individual is given by
\begin{equation}
P = G + \epsilon \: ,
\label{eq-pnaive}
\end{equation}
which says the phenotype of the focal individual is a linear function \textit{only} of its breeding value and a random component; thus, there is no effect of the partner's phenotype on the focal, which embodies the assumption we believe underlies \citet{Queller2011}'s analysis.  Plugging $P$ and $P'$ given in equations \eqref{eq-pprime} and \eqref{eq-pnaive} into equation \eqref{eq-queller-7} yields
\begin{align}
\beta _{W P \cdot P'}+\beta _{W P' \cdot P}\frac{\Cov\left[G,G'+\beta _{P' P}P+\epsilon '\right]}{\Cov[G,G+\epsilon ]}
& =\beta _{W P \cdot P'}+\beta _{W P' \cdot P}\left(\beta _{P' P}+\beta _{G' G}\right) \: ,
\end{align}
which can be written as
\begin{equation*}
-c+b(\rho +r)
\end{equation*}
in the notation of the main text and is equivalent to the additive combination of kin and kith components in equation (1) of \citet{Queller2011}.

To include bi-directional reciprocity, we follow the analysis presented by \citet{McGlothlin2010}; \citet{Akcay2012} obtain the same results and present a comparison of the relative merits of the current approach and the IGE approach. We start by setting 
\begin{equation}
P = G+\beta _{P P'}P'+\epsilon \: ,
\label{eq-p}
\end{equation}
which expresses the fact that $P'$ also has an effect on $P$. Before we can plug this definition of $P$ or the definition of $P'$ into equation (7) of \citet{Queller2011}, we must insert our definition of $P'$ into \eqref{eq-p} and our definition of $P$ into \eqref{eq-pprime} and solve the resulting equations; the solutions are
\begin{equation}
\begin{split}
P &=\frac{G+\epsilon +\beta _{P P'}\left(G'+\epsilon '\right)}{1-\beta _{P P'}\beta _{P' P}} \\
P' &=\frac{G'+\epsilon '+\beta _{P' P}(G+\epsilon )}{1-\beta _{P P'}\beta _{P' P}} \: .
\end{split}
\label{eq-p-recip}
\end{equation}
These solutions can be compared to the analogous solution in \citet{McGlothlin2010} (equation 6).  Plugging equations \eqref{eq-p-recip} into equation (7) of \citet{Queller2011} yields
\begin{align*}
\beta _{W P \cdot P'}+\beta _{W P' \cdot P}\frac{\Cov\left[G,\frac{G'+\epsilon '+\beta _{P' P}(G+\epsilon )}{1-\beta _{P P'}\beta _{P' P}}\right]}{\Cov\left[G,\frac{G+\epsilon +\beta _{P P'}\left(G'+\epsilon '\right)}{1-\beta _{P P'}\beta _{P' P}}\right]}
&=\beta _{W P \cdot P'}+\beta _{W P' \cdot P}\frac{\Cov\left[G,G'+\epsilon '+\beta _{P' P}(G+\epsilon )\right]}{\Cov\left[G,G+\epsilon +\beta _{P P'}\left(G'+\epsilon '\right)\right]}\\
&=\beta _{W P \cdot P'}+\beta _{W P' \cdot P}\frac{\Cov\left[G,G'\right]+\beta _{P' P}\Var[G]}{\Var[G]+\beta _{P P'}\Cov\left[G,G'\right]}\\
&=\beta _{W P \cdot P'}+\beta _{W P' \cdot P}\frac{\beta _{G' G}+\beta _{P' P}}{1+\beta _{P P'}\beta _{G' G}}
\end{align*}
Thus, the Hamilton's rule one gets from equation (7) of \citet{Queller2011} should be
\begin{equation*}
\beta _{W P \cdot P'}+\beta _{W P' \cdot P}\frac{\beta _{G' G}+\beta _{P' P}}{1+\beta _{P P'}\beta _{G' G}}>0
\end{equation*}
or
\begin{equation}
\left(1+\beta _{P P'}\beta _{G' G}\right)\beta _{W P \cdot P'}+\beta _{W P' \cdot P}\left(\beta _{G' G}+\beta _{P' P}\right)>0 \: .
\label{eq-queller-correct}
\end{equation}
Assuming that $\beta _{P' P}=\beta _{P P'}$, equation \eqref{eq-queller-correct} becomes in our notation
\begin{equation*}
-c(1+\rho  r)+b(r+\rho )>0 \: ,
\end{equation*}
which is exactly condition \eqref{incsym} from our model.  This is also exactly equivalent to equation (18) given by \citet{McGlothlin2010} for pairwise interactions once the appropriate mapping between $\rho$ and their IGE coefficient $\psi$ is made \citep[see Appendix A9 in ref][]{Akcay2012}.

\clearpage
\section{Interspecific mutualisms}
\label{sec:app-interspec}

The interaction between responsiveness and relatedness affects not only social behavior between individuals of the same species but it also shapes social behavior between species.
We can illustrate this by applying the population genetic model with continuous traits (Section \ref{sec:popgen} in the main text and Appendix \ref{sec:app-class})
to an interaction between a population of host individuals that each interact with a group of $n$ symbionts. Label the investment of the host species ($H$) into the $i$th symbiont by $x_i$ and the investment of the $i$th symbiont ($S_i$) into the host by $y_i$. The payoffs (i.e., individual fertility as a result of investments) are denoted $u_H(x_1,\cdots,x_n,y_1,\cdots,y_n)$ for the host and $u_{S_i}(x_i, y_i)$ for the $i$th symbiont (i.e., we assume that a symbiont's fertility is only affected by the investment it makes into and it receives from the host). 

Suppose further that the investments of both the hosts and symbionts are plastic and can change in response to each other. As a result, all of the investment variables are joint functions of the genotypes of the individual hosts and symbionts. Following the methodology in Section \ref{sec:popgen} \citep[see also][]{Akcay2009,Akcay2012}, we label the host's and symbionts' genetically determined phenotypes $p_H$ and $p_S$, respectively. Recall that phenotype in this context refers to a physiological mechanism (neurological, biochemical, etc) that alters investment levels, and that phenotypes are directly genetically determined rather investment levels. To find the increase condition for genotypes that result in higher investments in the mutualism, we apply equation \eqref{eq:1order} once for each species while treating the partner of the other species as a non-reproductive class.

Calculating the increase condition requires the total derivative of the fertilities ($u_H$ and $u_S$) with respect to the host and symbiont phenotypes. For example, the change is host fertility with respect to a change in the phenotype of symbiont $j$ is:
\begin{align}
\frac{d u_H}{d p_{Sj}}&=\sum_{i}\frac{\partial u_H}{\partial x_i}\frac{d x_i}{dp_{Sj}}+\sum_{i}\frac{\partial u_H}{\partial y_i}\frac{d y_i}{dp_{Sj}}\nonumber\\
& \propto n \left[- c_H + \rho_{HS} b_H\right]\:,
\end{align}
where $c_H\equiv-\frac{\partial u_H}{\partial x_i}$, $b_H \equiv\frac{\partial u_H}{\partial y_i}$, and $\rho_{HS}\equiv\frac{d y}{dp_{Sj}}/\frac{dx}{dp_{Sj}}$. The latter quantity denotes the response of a symbiont's investment to the change of the host's investment into that symbiont \citep{Foster2006,Akcay2011,Akcay2014}. Calculating the other total derivatives in the same way, we obtain the (partial) conditions for the ESS phenotypes in hosts and symbionts, respectively, as
\begin{align}
 \frac{b_H}{c_H}&=\frac{1 + r_{HS} \left(\rho_{S_{\bullet} H_{\bullet}} + (n-1) \rho_{S_{\bullet} H_{\circ}} \right)}{\rho_{HS}+r_{HS} \left(1+(n-1)\rho_{SS} \right)} \label{hostinc}\\
 \frac{b_S}{c_S}&=\frac{1 + r_{SH} \rho_{HS} + (n-1) r_{SS} \rho_{SS}}{\rho_{S_{\bullet} H_{\bullet}} + (n-1) r_{SS} \rho_{S_{\bullet} H_{\circ}}+r_{SH}} \: , \label{symbinc}
\end{align}
where $r_{SS}$ is the relatedness coefficient for the symbionts. The coefficients $r_{HS}$ and $r_{SH}$ measure the genetic covariance between higher investment level phenotypes of the symbionts and the host (relative to host and symbiont genetic variance, respectively) and are analogous to relatedness coefficients \citep{Frank1994}. The conditions above are only partial ESS conditions, since the between species genetic association coefficients are dependent on the host and symbiont phenotypes and their fertility effects. To generate the full ESS conditions, we would need a set of equations that describe the population dynamics of host and symbionts as a function of their investment phenotypes and find the equilibrium values for the genetic associations in the structured population, as is done by \citet{Frank1994}. Finally, the coefficient $\rho_{S_{\bullet} H_{\bullet}}$ denotes how much increased investment of a focal symbiont causes the host to increase its investment into the same symbiont,  $\rho_{S_{\bullet} H_{\circ}}$ measures how the investment in another symbiont by the host responds to a focal symbiont's investment, and $\rho_{SS}$ measures the response of symbionts to each other. 

Similar to the intraspecific condition for the increase of cooperative phenotypes (equation \ref{eq:1order:bc}), conditions \eqref{hostinc} and \eqref{symbinc} contain both multiplicative and additive interactions between the different relatedness and responsiveness coefficients.  Such interactions have not been fully appreciated before \citep[e.g.,][]{Foster2006} and have important effects on the level of investments that evolve in the host and symbionts.  For example, hosts and symbionts that can respond plastically to changes in the investment levels of one another result in higher levels of investment by both host and symbiont and this increase is mediated by an interaction between relatedness and responsiveness.


\clearpage

\clearpage

\clearpage

\clearpage

\end{document}